\documentclass[twocolumn]{aastex63}
\usepackage{graphicx}
\usepackage{epsfig}
\usepackage{color}
\usepackage{txfonts}
\usepackage{natbib,twoopt}
\usepackage{soul}

\newcommand{\Msun}{\mbox{M$_{\odot}$}}
\newcommand{\Myr}{\mbox{M$_{\odot}$ yr$^{-1}$}}
\newcommand{\CIV}{C{\sc iv}}
\newcommand{\OVI}{O{\sc vi}}
\newcommand{\Lya}{Ly$\alpha$}
\newcommand{\Ha}{H$\alpha$}
\newcommand{\Hb}{H$\beta$}
\newcommand{\HI}{\mbox{H\,{\sc i}}}
\newcommand{\HII}{\mbox{H\,{\sc ii}}}
\newcommand{\Hmol}{\mbox{H$_{\rm 2}$}}

\newcommand{\CII}{\mbox{[C\,{\sc ii}]}}
\newcommand{\NII}{\mbox{N\,{\sc ii}}}
\newcommand{\OI}{\mbox{[O\,{\sc i}]}}
\newcommand{\nH}{$n_{\rm H}$}
\newcommand{\mH}{$m_{\rm H}$}
\newcommand{\Wm}{W~m$^{-2}$}

\def\approxlt{\lower.2em\hbox{$\buildrel < \over \sim$}}
\def\approxgt{\lower.2em\hbox{$\buildrel > \over \sim$}}

\def\gtrsim{\mathrel{\hbox{\rlap{\hbox{\lower4pt\hbox{$\sim$}}}\hbox{$>$}}}}
\newcommand{\kms}{\mbox{{\rm km}\,{\rm s}$^{-1}$}}

\def\lesssim{\mathrel{\hbox{\rlap{\hbox{\lower4pt\hbox{$\sim$}}}\hbox{$<$}}}}

\def\la{\mathrel{\hbox{\rlap{\hbox{\lower4pt\hbox{$\sim$}}}\hbox{$<$}}}}
\def\ga{\mathrel{\hbox{\rlap{\hbox{\lower4pt\hbox{$\sim$}}}\hbox{$>$}}}}

\begin{document}

\submitjournal{ApJ}

\title{Extremely broad \Lya\ line emission from the molecular intra-group medium in Stephan's Quintet: evidence for a turbulent cascade in a highly clumpy multi-phase medium?}



\submitjournal{ApJ}
\author{P. Guillard}
\affiliation{Sorbonne Universit\'{e}, CNRS, UMR 7095, Institut d'Astrophysique de Paris, 98bis bd Arago, 75014 Paris, France}
\affiliation{Institut Universitaire de France, Minist{\`e}re de l'Enseignement Sup{\'e}rieur et de la Recherche, 1 rue Descartes, 75231 Paris Cedex 05, France}
\author{P. N Appleton}
\affiliation{Caltech/IPAC, MC 6-313, 1200 E. California Blvd., Pasadena, CA 91125, USA}
\author{F. Boulanger}
\affiliation{Laboratoire de Physique de l'Ecole Normale Sup\'{e}rieure, ENS, Universit\'{e} PSL, CNRS, Sorbonne Universit\'{e}, Universit\'{e} de Paris, 75005 Paris, France}
\author{J. M. Shull}
\affiliation{CASA, Astrophysical and Planetary Sciences Dept., University of Colorado, UCB-389, Boulder, CO 80309, USA}
\author{M.~D. Lehnert}
\affiliation{Sorbonne Universit\'{e}, CNRS, UMR 7095, Institut d'Astrophysique de Paris, 98bis bd Arago, 75014 Paris, France}
\author{G. Pineau des Forets}
\affiliation{Observatoire de Paris, PSL University, Sorbonne Universit\'{e}, LERMA, 75014 Paris, France}
\affiliation{Universit\'{e} Paris Saclay, CNRS, Institut d'Astrophysique Spatiale, 91405 Orsay, France}
\author{E. Falgarone}
\affiliation{Laboratoire de Physique de l'Ecole Normale Sup\'{e}rieure, ENS, Universit\'{e} PSL, CNRS, Sorbonne Universit\'{e}, Universit\'{e} de Paris, 75005 Paris, France}
\author{M.E. Cluver}
\affiliation{Centre for Astrophysics and Supercomputing, Swinburne University of Technology, John Street, Hawthorn 3122, Victoria, Australia}
\affiliation{Department of Physics and Astronomy, University of the Western Cape, Robert Sobukwe Road, Bellville, South Africa}
\author{C.K. Xu}
\affiliation{IPAC, Caltech, MC 100-22, 1200 E. California Blvd., Pasadena, CA 91125, USA}
\affiliation{Chinese Academy of Sciences South America Center for Astronomy, National Astronomical Observatories, CAS, Beijing 100101, China}
\author{S.C. Gallagher}
\affiliation{Department of Physics and Astronomy, University of Western Ontario, London, ON N6A 3K7, Canada}
\author{P.A. Duc}
\affiliation{Universit\'{e} de Strasbourg, CNRS, Observatoire astronomique de Strasbourg, UMR 7550, F-67000 Strasbourg, France}

\shorttitle{Broad \Lya\ line emission from the Stephan's Quintet shock}
\shortauthors{Guillard, Appleton, Boulanger, Shull et al.}

\received{February, 2021}
\revised{September 20th, 2021}
\accepted{October 18th, 2021}

\begin{abstract} 
We present \textit{Hubble Space Telescope} Cosmic Origin Spectrograph (COS) UV line spectroscopy and integral-field unit (IFU) observations of the intra-group medium in Stephan's Quintet (SQ). 
SQ hosts a 30 kpc long shocked ridge triggered by a galaxy collision at a relative velocity of 1000~\kms, where large amounts of molecular gas coexist with a hot, X-ray emitting, plasma. COS spectroscopy at five positions sampling the diverse environments of the SQ intra-group medium reveals very broad ($\approx 2000$~\kms) \Lya\ line emission with complex line shapes. 
The \Lya\ line profiles are similar to or much broader than those of \Hb, \CII157.7$\mu$m and CO~(1-0) emission. The extreme breadth of the \Lya\ emission, compared with \Hb, implies resonance scattering within the observed structure.
Scattering indicates that the neutral gas of the intra-group medium is clumpy, with a significant surface covering factor.
We observe significant variations in the \Lya/\Hb\ flux ratio between positions and velocity components.  From the mean line ratio averaged over positions and velocities, we estimate the effective escape fraction of \Lya\ photons to be $\approx 10 - 30$\%. 
 Remarkably, over more than four orders of magnitude in temperature, the powers radiated by  X-rays, \Lya, \Hmol, \CII\ are comparable within a factor of a few, assuming that the ratio of the \Lya\ to \Hmol\ fluxes over the whole shocked intra-group medium stay in line with those observed at those five positions. Both shocks and mixing layers could contribute to the energy dissipation associated with a turbulent energy cascade. Our results may be relevant for the cooling of gas at high redshifts, where the metal content is lower than in this local system, and a high amplitude of turbulence is more common. 
\end{abstract}


\section{Introduction}

Galaxy interactions are important phases of their evolution, often involving high-speed shocks and dissipation or large amounts of kinetic energy. Many of these interactions are observed in infrared (IR) and visible light to trigger bursts of star-formation. The dissipation of kinetic energy affects the gas cooling and how, when, and where star formation proceeds. To make headway in our understanding of the conversion of molecular gas to stars, it is crucial to determine the mechanism and rate of gas cooling. 

Stephan's Quintet (HCG~92, hereafter SQ) is a compact group of five interacting galaxies \citep{Arp1973} with a complex dynamical history, involving multiple galaxy collisions \citep{Moles1997, Renaud2010, Hwang2012, Duc2018}. It is an ideal laboratory for the study of galaxy interactions and their impact on the physical state and energetics of the gas, especially the dissipation of merger-driven turbulence isolated against the dark sky \citep{Guillard2009}. It is one of the few extragalactic sources where one can spatially separate star forming regions from shocked gas. When excluding the large foreground dwarf galaxy NGC~7320, the main group is dominated by four large massive galaxies. Three of them, NGC~7317, NGC~7318a and NGC~7319, have heliocentric radial velocities in the range V$_{helio} = 6599-6747$~\kms, and together they define the main barycentric velocity of the group (at around 6600~\kms). A fourth galaxy, NGC~7318b (V$_{helio}$ = 5774~\kms, is often described as an intruder galaxy because it appears to be colliding into NGC 7319's tidal tail in the intra-group medium at a relative velocity of $\approx \! 1000$~\kms \citep[see Fig~\ref{fig:SQ_contours_COS_on_WFC3Ha} and][]{Xu2003}. 

The collision of NGC~7318b with the gas in the intra-group medium is believed to be responsible for a striking feature of the group, namely a galaxy-wide shock ($\approx 15 \times 35$~kpc$^{2}$; Fig.~\ref{fig:SQ_contours_COS_on_WFC3Ha}) seen at many wavelengths. A ridge of X-ray \citep{OSullivan2009} and radio synchrotron \citep{Allen1972} emission from the hot ($6 \times 10^6\,$K) post-shock plasma is associated with the group-wide shock. 
Surprisingly, \textit{Spitzer} IRS spectroscopy revealed that the mid-IR spectrum in the intra-group medium  is dominated by the rotational lines of  $\rm H_2$ (see red contours on Fig.~\ref{fig:SQ_contours_COS_on_WFC3Ha}), with very weak dust emission from star formation  \citep{Appleton2006, Cluver2010}. Over the shock region, no \HI\ emission is detected, but optical line emission from ionized gas is observed at the velocity of NGC 7319's \HI\  tidal tail \citep{Sulentic2001}.

The weakness of the mid-IR tracers of star-formation (dust features and ionized gas lines) relative to the H$_2$ lines suggests that, despite the large H$_2$ mass estimated to be $\approx 5 \times 10^9 \,$\Msun\ \citep{Guillard2012a}, the star formation rate is on average very low in the shock \citep[$<0.07\,$\Myr,][]{Cluver2010}. This is a factor 40 below the star formation rate expected from the Schmidt-Kennicutt relation \citep{Guillard2012a}. Deep HST/WFC3 \Ha\ imaging of SQ reveals many compact \Ha\ knots and filaments spread over the shock, as well as a diffuse, underlying component \citep{Gallagher2001}.
Extensive Gemini optical spectroscopy of 50 \Ha\ knots in the group \citep{Konstantopoulos2014} shows that both photo-ionization and shock excitation are present  \citep[see also][for earlier long-slit optical spectroscopy]{Xu2003}, a result that is also confirmed by IFU spectroscopy \citep{DuartePuertas2019}. Some knots are star clusters \citep[masses from $10^4$ to a few $10^5\,$\Msun,][]{Gallagher2001, Fedotov2011}, while others show very strong [O{\sc iii}]$\lambda \, 5007\,$\AA~and  [O{\sc i}]$\lambda \, 6300\,$\AA~emission, and very broad (FWHM up to $\approx 700\,$km~s$^{-1}$), complex optical emission line profiles, consistent with pure shock excitation \citep{Konstantopoulos2014}.  

Spitzer and Herschel have shown that the \Hmol, \CII$\lambda158\mu$m and \OI$\lambda63\mu$m IR lines are important coolants of the shocked gas \citep{Appleton2013, Appleton2017}. \citet{Guillard2009} proposed a model in which this line emission is powered by the dissipation of kinetic energy, through a turbulent cascade of energy, from the galaxy-wide shock down to low-velocity shocks within molecular gas.  GALEX imaging of SQ \citep{Xu2005} shows that most of the UV emission in the intra-group medium ridge is associated with an arc-like structure to the South-East of the intruder NGC~7318, which spatially correlates with a chain of \Ha\ knots. The broad-band UV emission at the center of the shock, where the \Hmol\ emission peaks, is extended and corresponds to a radiation field of average intensity $G_{UV} = 1.4$~Habing \citep{Guillard2010}. 

This paper complements the rich array of SQ observations by presenting UV spectra obtained with HST COS, focusing on the \Lya\ line emission. We assess the origin of the \Lya\ photons and their contribution to the energy budget of the SQ galaxy collision. The  comparison of the line profiles with those of CO, \CII\ and \Hb\ provide insight on the structure of the multiphase ISM in SQ.

We assumed a distance to Stephan's Quintet of 94~Mpc for H$_0$ = 70 km s$^{-1}$ Mpc$^{-1}$, and a group systemic heliocentric velocity of 6600 km s$^{-1}$.

\section{Observations and data reduction}

\subsection{Observed COS line-of-sights}

We chose to perform deep COS spectroscopy of 5 regions in the intra-group medium of SQ, as shown on Fig.~\ref{fig:SQ_contours_COS_on_WFC3Ha}, for which we have very high signal-to-noise optical spectra from Gemini. The COS pointings are also associated with CO(1-0) line emission detected with the NOEMA interferometer, as well as the 30m single-dish for positions 1, 2, 5 and 7. The optical line properties derived from Gemini, the slit positions associated with the COS beams, and the CO(1-0) line properties are gathered in Table~\ref{table:optical_lines}. Our five pointings were also chosen to probe diverse environments. As the background HST WFC3 image in Fig.~\ref{fig:SQ_contours_COS_on_WFC3Ha} shows, positions 1, 2 and 7 are associated with bright \Ha\ knots in the intra-group medium ridge, while positions 3 and 5 are not. Positions 1, 2 and 5 are associated with very broad optical line emission, with line ratios consistent with pure shock excitation \citep[see Fig.~10 in][]{Konstantopoulos2014}. Position 5 lies in the so-called \textit{bridge} between NGC~7319 and the ridge, and shows faint, diffuse \Ha\ emission filling up the COS beam, associated with NGC~7319's tidal tail. Position 3 points towards a region of the ridge devoid of compact \Ha\ emission. The associated Gemini slit is slightly offset from the COS beam and hits a bright \Ha\ knot, whose excitation is consistent with that of an \HII\ region, so comparison between COS and Gemini data at this position may not be meaningful. Position 7 is located in the SQ-A starburst region in the Northern region of the ridge, and is one of the two brightest star-forming regions in the intra-group medium of SQ. This region is known to combine photo-ionization from star formation and shocks. Indeed, for instance, optical Gemini spectra show strong variation in the \Ha/\NII\ line ratio  as a function of gas velocity. In this paper, to characterize the dissipation of turbulence in the intra-group medium environment of SQ, we focus on regions known to be shock-dominated.

\subsection{Reduction of COS spectroscopic data}

SQ was observed with the medium-resolution far-UV G130M (\Lya) and G160M (\CIV) gratings of \textit{HST}-COS on 2014 November 9 for a total of 15 orbits. Descriptions of the COS instrument and on-orbit performance characteristics can be found in \citep{Green2012, Osterman2011}, as well as in the COS Instrument Handbook.
In order to achieve continuous  spectral coverage across the G130M bandpass (1135-1440~\AA) and to minimise fixed-pattern noise, we
 made observations at two central wavelength settings (1291~\AA\ and 1300~\AA) with four focal-plane offset  locations in each grating setting (i.e. FP-POS = 1, 2, 3, 4).
 This combination of grating settings ensures the
 highest signal-to-noise observations at the shortest wavelengths available to the G130M mode at a resolving power of $R = \lambda / \Delta \lambda \approx 16,000 $ (about 17 \kms velocity resolution for a point source). 
For every sightline, COS observations yielded a continuous spectrum spanning $\lambda \approx 1150-
1800$~\AA. The calibration of the wavelength scale with updated dispersion solutions ensures a velocity accuracy of 7~\kms. The exposure times were chosen to achieve a signal-to-noise ratio (S/N) of 7--15 per resolution element (7 pixels, $\approx 0.07 \AA$, ${\rm FWHM} = 15\ \kms$ for a point source) at $\lambda \approx 1300$~\AA, depending on the velocity range. We note that the spectral resolution could be significantly lower for extended sources, as low as $R = 1450$ for G130M if the emission uniformly fills the COS beam. In particular, this may be the case for positions 3 and 5, which do not show an associated compact \Ha\ source within the COS aperture. 
The target positions, grism configurations and individual exposure times are presented in Table~1. We plot the spectra binned to the standard 7 pixels.

We started our data reduction from the x1d.fits files produced by the COS calibration  pipeline, {\it CalCOS}\footnote{see HST COS Instrument Handbook for more details:
\url{http://www.stsci.edu/hst/cos/documents/handbooks/current/cos_cover.html}.}, downloaded from MAST. To assess the contamination of the spectra by geocoronal airglow lines, we have filtered the data in time in order to produce spectra utilizing only data taken during orbital night time. To do that, we used the {\it TimeFilter} routine from the COSTOOLS package\footnote{\url{https://github.com/spacetelescope/costools}}  to select night time data in the COS corrtag files, and then re-extracted the data into x1d spectra. The comparison between the night-only and full datasets did not yield significant improvement because the redshift of the source (0.02) shifts the \Lya\ line  to 1240~\AA, in between the geocoronal \Lya\ and \OI\ lines.  The data have then been aligned and coadded with the COADD\_X1D.pro V3.3 IDL routine provided by STScI\footnote{available at \url{http://casa.colorado.edu/~danforth/science/cos/costools}}. To take into account the strong wings of the non-Gaussian, Line Spread Function (LSF) of COS, we have used the COS\_LSF.pro IDL routine to produce a LSF model at the nearest tabulated wavelength value. The COS exposures for each regions were deconvolved using the LSF appropriate for corresponding grating setting and for the central wavelengths of the \Lya\ and \CIV\ emission lines.
The absolute flux calibration steps are described in detail in the COS data handbook\footnote{\url{https://hst-docs.stsci.edu/cosdhb/chapter-3-cos-calibration}}, and are expected to of the order of $\pm 5$\% for G130M and G160M (errors are dominated by fixed pattern noise in the UV detectors and uncertainties in the time-dependent sensitivity correction). 
 
\subsection{Visible light IFU spectroscopy}
We also present Integral Field Unit (IFU) observations made with the George and Cynthia Mitchell Spectrograph (hereafter GCMS, formerly known as VIRUS-P) mounted on the 2.7~m Harlan J. Smith Telescope at McDonald Observatory \citep{Hill2008,Blanc2013}. The IFU uses a 246 fiber bundle, with each fiber covering $4\farcs16$ on the sky, which makes it sensitive to faint extended emission.  We used a 3-point dither pattern to completely cover the 2.82~arcmin$^2$  field of view, and to fill in gaps between the fibers. Observations were obtained on 1 Oct 2011 with an integration time of one hour at each of the three dither positions. We used the VP2 blue grating, which has a spectral resolution of $\sim1.6$ \AA\ ($\sim$ 100 \kms) and covers the wavelength range $\sim4700$  to $\sim5350$~\AA.  The data reduction was performed using the VACCINE software package \citep{Adams2011}, and these data were further flux calibrated using a bootstrapping method  which compares each fiber response to a calibrated SDSS b-image of the galaxy \citep[e.g.][]{Joshi2019}. Wavelength calibration was performed using lamp spectra obtained at the beginning of the observing run.

\begin{figure*}
\includegraphics[width=\textwidth]{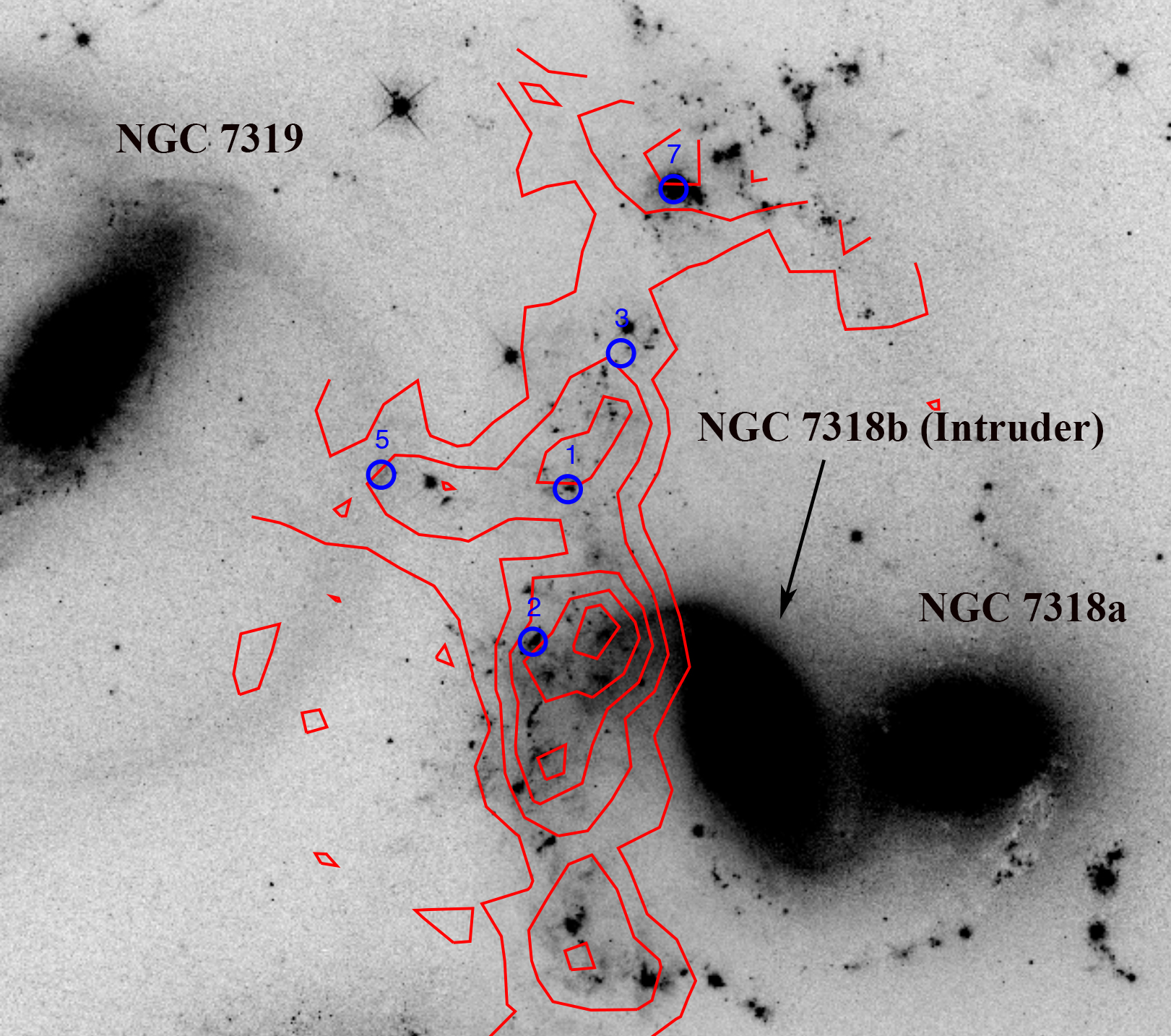}
\caption{\textit{HST} Wide Field Camera 3 F665N image of the inner SQ group with \textit{Spitzer} IRS H$_2$ (0-0)~S(3) 9.6$\mu$m  line flux contours in red from \citet{Cluver2010}. This emission from warm molecular hydrogen highlights the North-South shocked ridge of the intra-group medium, as well as an extension towards NGC~7319, called ''the bridge''. The blue circles are the COS apertures (2.5" in diameter) corresponding to 1.1~kpc at a distance $D = 94$~Mpc. The positions, observing parameters and integration times are listed in Table~1.}
\label{fig:SQ_contours_COS_on_WFC3Ha}
\end{figure*}


\begin{deluxetable}{cccccc}
\tablecolumns{6}
\tablecaption{COS observation log: MAST archive name, positions, COS gratings, central wavelength and individual exposure times for each of the 5 positions observed.}
\tablehead{
\colhead{COS target} &	\colhead{RA} & \colhead{Dec} & \colhead{Grating} & \colhead{$\lambda _{cen}$} & \colhead{Exp. time} \\
\colhead{(MAST)}  & 	\colhead{(J2000)}  & \colhead{(J2000)}  &   &	\colhead{[\AA]}  & \colhead{[seconds]} 
}
\startdata
HCG92-1	& 22 35 59.765 & +33 58 21.33 &	G130M & 1096 & 1287.872	\\
        &              &              & G130M & 1222 & 1437.696	\\
        &              &              & G160M & 1611 & 3180.512	\\        
        &              &              & G160M & 1623 & 3673.536 \\	
\hline
HCG92-2	& 22 36 00.032 & +33 58 06.75 & G130M & 1096 & 1287.936	\\
        &              &              & G130M & 1222 & 1437.760	\\
        &              &              & G160M & 1611 & 3180.544	\\        
        &              &              & G160M & 1623 & 3673.568 \\
\hline
HCG92-3	& 22 35 59.439 & +33 58 34.80 & G130M & 1096 & 1429.056	\\
        &              &              & G130M & 1222 & 1447.840	\\
        &              &              & G160M & 1611 & 3180.576	\\        
        &              &              & G160M & 1623 & 3673.568	\\
\hline 
HCG92-5	& 22 36 01.222 & +33 58 22.74 & G130M & 1096 & 1366.016	\\
        &              &              & G130M & 1222 & 1359.552	\\
        &              &              & G160M & 1611 & 3180.640	\\        
        &              &              & G160M & 1623 & 3671.584	\\
\hline 
HCG92-7	& 22 35 58.953 & +33 58 49.96 & G130M & 1096 & 600.000	\\
        &              &              & G130M & 1222 & 599.712	\\
        &              &              & G160M & 1611 & 1102.656	\\        
        &              &              & G160M & 1623 & 1303.712	
\enddata
\label{table:COS_obslog}
\end{deluxetable}

\begin{deluxetable*}{lrrrrr}
\tablecolumns{6}
\tablecaption{Observed \Lya, \Hb\ and H$_2$ line fluxes and line ratios for the COS lines of sight.}
\tablehead{
\colhead{COS target}  & \colhead{$F_{Ly\alpha}$: \Lya\ Flux\tablenotemark{a}} & \colhead{$F_{H\beta}$: \Hb\ flux\tablenotemark{b}} & \colhead{F$_{H2}$: H$_2$ flux \tablenotemark{c}} & \colhead{$F_{Ly\alpha}$/$F_{H\beta}$} & \colhead{$F_{Ly\alpha}$/F$_{H2}$}  \\
		       & \colhead{[$10^{-18}\,$\Wm]}    &	\colhead{[$10^{-19}\,$\Wm]}  			     &      \colhead{[$10^{-18}\,$\Wm]}   &  &                     
}
\startdata
HCG92-1	(ridge)    & $5.3 \pm 0.5$ & $1.7 \pm 0.2$ &	$6.7 \pm 0.8$ & $30.9 \pm 3.4$ & $0.77 \pm 0.12$	\\
HCG92-2	(ridge)    & $5.2 \pm 0.5$ & $2.4 \pm 0.2$ &	$7.3 \pm 0.9$ & $22.0 \pm 2.5$ & $0.71 \pm 0.11$	\\
HCG92-3	(ridge)    & $1.1 \pm 0.1$ & $1.6 \pm 0.3$ &	$5.8 \pm 0.7$ &  $7.0 \pm 1.2$ & $0.19 \pm 0.03$  \\
HCG92-5	(bridge)    & $0.4 \pm 0.1$ & $0.8 \pm 0.1$ &	$5.5 \pm 0.6$ &  $4.7 \pm 0.9$ & $0.07 \pm 0.02$  \\
HCG92-7	(SQ-A)    & $1.3 \pm 0.1$ & $6.7 \pm 0.1$ &	$7.0 \pm 0.8$ &  $1.8 \pm 0.2$ & $0.17 \pm 0.02$ 	\\
all (stacked)	&$13.3 \pm 0.2$ &$13.2 \pm 0.4$ &$32.3 \pm 1.7$ & $10.1 \pm 1.4$ & $0.41 \pm 0.03$  \\
\enddata
\tablenotetext{a}{Uncorrected for \Lya\ absorption. Some of the profiles show evidence of absorption (see text). }
\tablenotetext{b}{Sum of the observed line flux estimated on a $4\times 4$~arcsec$^2$ square aperture from the Mitchell Spectrograph IFU data, scaled to the circular COS aperture of 2.5" in diameter assuming constant surface brightness.}
\tablenotetext{c}{Sum of the (0-0)~S(0) + (0-0)~S(1) + (0-0)~S(2) + (0-0)~S(3) pure rotational H$_2$ line fluxes derived from extractions of $5.5\times5.5$~~arcsec$^{2}$ from \textit{Spitzer} IRS data \citep{Appleton2017}, scaled down to the circular COS aperture of 2.5" in diameter assuming constant surface brightness.}
\label{table:fluxes}
\end{deluxetable*}

\begin{deluxetable}{lrrr}
\tablecolumns{4}
\tablecaption{Observed \Lya\ line kinematical properties for the COS lines of sight.}
\tablehead{
\colhead{COS target}  & \colhead{$W_{10}$(\Lya)\tablenotemark{a}} & \colhead{$W_{20}$(\Lya)\tablenotemark{a}} & \colhead{$W_{50}$(\Lya)\tablenotemark{a}} \\
		   & \colhead{[\kms]}  & \colhead{[\kms]} & \colhead{[\kms]} 
}
\startdata
HCG92-1	      &  $1910 \pm 40$ &  $1690 \pm 20$ & $1290 \pm 40$ \\
HCG92-2	      &  $2120 \pm 40$ &  $1080 \pm 20$ &  $690 \pm 20$ \\
HCG92-3	      &  $2000 \pm 40$ &  $1770 \pm 20$ &  $570 \pm 20$ \\
HCG92-5	      &  $1120 \pm 40$ &  $930  \pm 40$ &  $350 \pm 20$ \\
HCG92-7	      &  $1480 \pm 40$ &  $530  \pm 40$ &  $220 \pm 20$ \\
all (stacked) &  $2650 \pm 40$ 	& $1820  \pm 20$ &  $595 \pm 15$
\enddata
\tablenotetext{a}{Widths at 10\%, 20\% and 50\% of maximum flux.}
\tablecomments{$W_{10}$ has been computed after smoothing the spectra to 80~\kms\ to increase the SNR in the line wings, and is close to the Full Width at Zero Intensity of the line. In case of the presence of several velocity components, only the widest component is listed.}
\label{table:Lya_widths}
\end{deluxetable}

\begin{deluxetable*}{l l l l l l l l l l}
\tablecolumns{9}
\tablecaption{CO~(1-0) line properties extracted at the positions of the COS apertures and optical line properties derived from Gemini spectroscopy at the nearest position of the HST COS apertures.}
\tablehead{
\colhead{COS target} & \colhead{$I_{\rm CO}$\tablenotemark{a}} & \colhead{$v_{\rm CO}$\tablenotemark{a}} & \colhead{$\sigma _{\rm CO}$\tablenotemark{a}} & \colhead{Gemini Id\tablenotemark{b}} & \colhead{ F(\Ha) \tablenotemark{c}} & \colhead{[N{\sc ii}]/\Ha \tablenotemark{c}}  & \colhead{\OI/\Ha \tablenotemark{c}} & \colhead{\Hb/\Ha \tablenotemark{c}}  \\
         &        \colhead{[Jy~\kms]}   &        \colhead{[\kms]}  	&	\colhead{[\kms]}     &  	&	 \colhead{ [$10^{-18}\,$\Wm]}		&	& 	
}
\startdata
HCG92-1 (ridge) & $2.6\pm 0.2$ & 6084.9 & 28.8 & 15053$^*$ & $0.0774 \pm 0.0058 $ & $ 0.168 \pm 0.001$ & $ 0.251 \pm 0.002 $ & $ 0.320 \pm 0.002 $  \\
HCG92-2 (ridge) & $1.8 \pm 0.2$ & 6080.1 & 30.6 & 12002 & $0.2553 \pm 0.0180 $ & $ 0.141 \pm 0.001$ & $ 0.203 \pm 0.002 $ & $ 0.145 \pm 0.003 $  \\
HCG92-3 (ridge) & $2.2 \pm 0.3$ & 6260.0 & 35.7 & 14006$^*$ & $0.0639 \pm 0.0016 $ & $ 0.129 \pm 0.001$ & $ 0.025 \pm 0.001 $ & $ 0.241 \pm 0.001 $    \\
HCG92-5 (bridge) & $1.1 \pm 0.3$ & 6350.0 & 67.8 & 20001 & $0.0276 \pm 0.0007 $ & $ 0.210 \pm 0.002$ & $ 0.217 \pm 0.001 $ & $ 0.460 \pm 0.002 $  \\
HCG92-7 (SQ-A)& $4.3 \pm 0.3$ & 6736.8 & 19.8 & 30109 & $1.0383 \pm 0.0032 $ & $ 0.051 \pm 0.001$ & $ 0.217 \pm 0.001 $ & $ 0.197 \pm 0.002 $ 	
\enddata
\tablenotetext{a}{Parameters estimated from the CO~(1-0) spectrum extracted from the IRAM NOEMA interferometer data (Guillard et al. in prep.) on a beam size $4.2 \times 3.9$~arcsec$^{2}$: integrated intensity, central velocity and velocity dispersion.}
\tablenotetext{b}{Closest Gemini slit ID from \citet{Konstantopoulos2014}. The star indicates when the Gemini slit is slightly offset from the COS aperture.
}
\tablenotetext{c}{\Ha\ flux and optical line ratios from \citet{Konstantopoulos2014}. }
\label{table:optical_lines}
\end{deluxetable*}

\begin{figure*}[!h]
\includegraphics[width=0.49\textwidth]{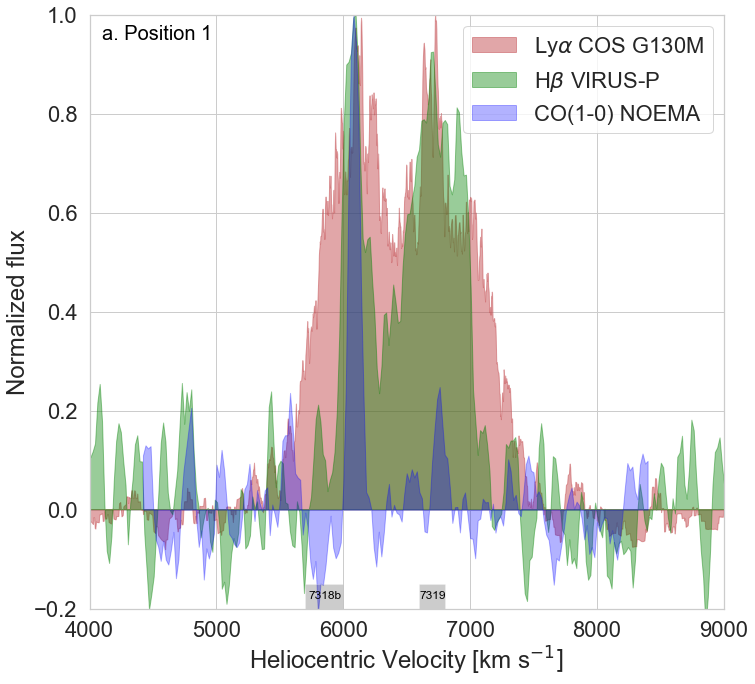}
\includegraphics[width=0.49\textwidth]{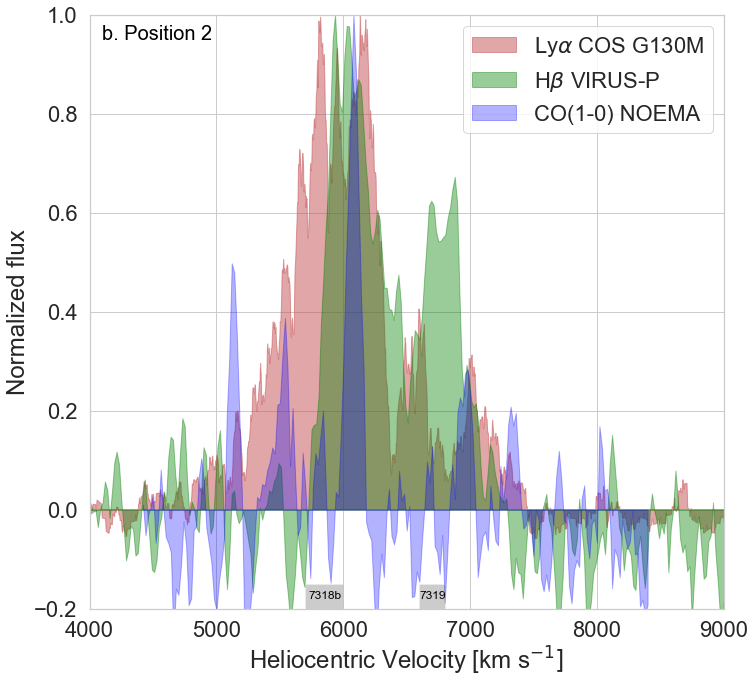}
\includegraphics[width=0.49\textwidth]{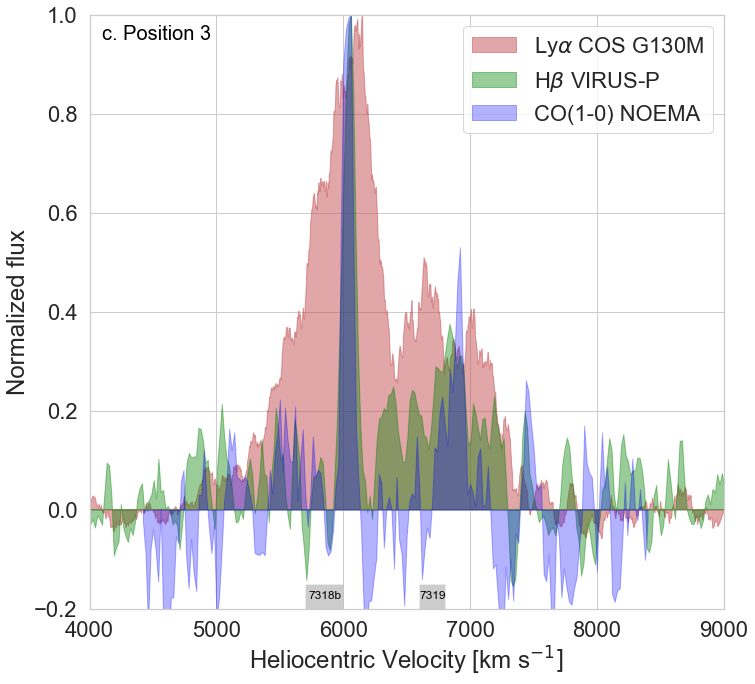}
\includegraphics[width=0.49\textwidth]{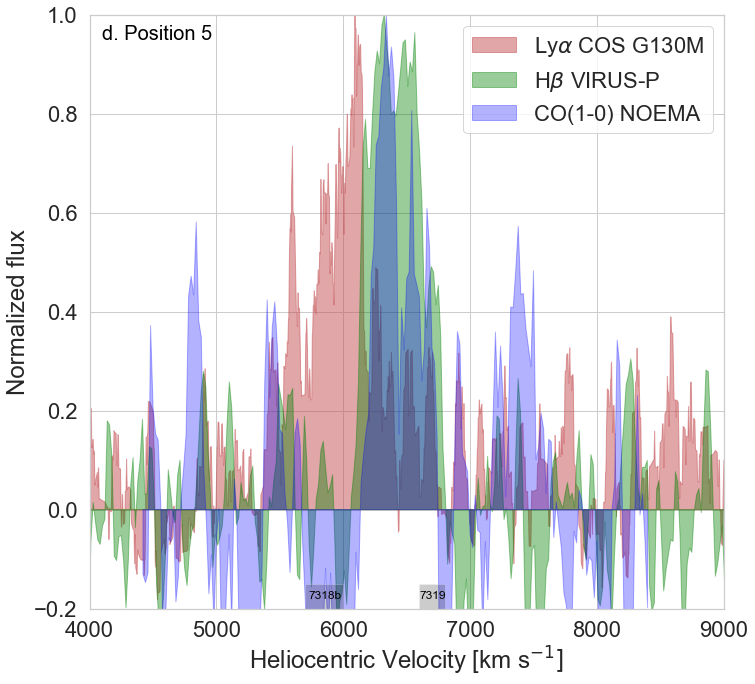}
\includegraphics[width=0.49\textwidth]{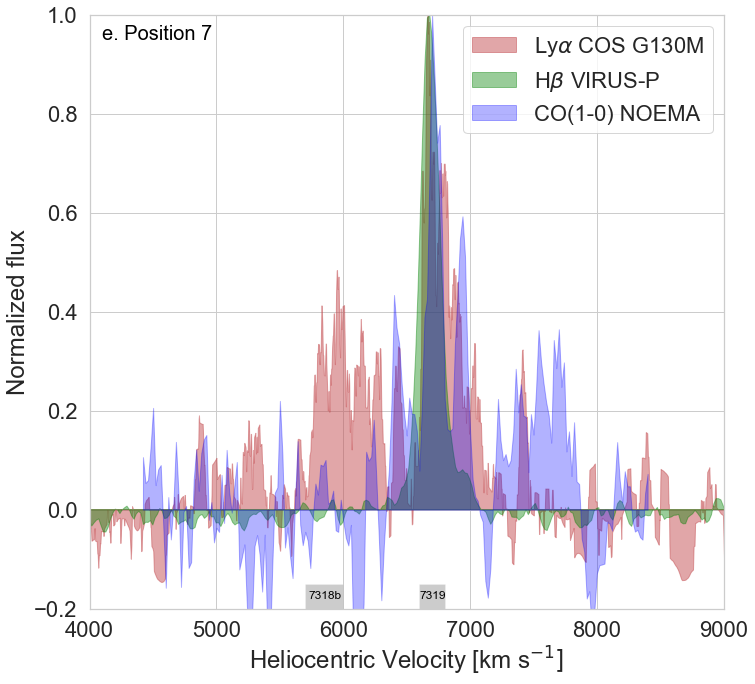}
\caption{Comparison of \Lya\ (red), \Hb\ (green), and CO~(1-0) (blue) line profiles for the 5 positions observed. The CO~(1-0) PdBI spectra are extracted at the COS positions over an ellipsoidal beam of $4.3 \times 3.5$ arcsec$^2$, P.A = 100~degrees. The \Lya\, CO(1-0), \Hb\ spectra have respectively a velocity resolution of 20, 30, and 100~\kms. }
\label{fig:hcg92_COS_CO}
\end{figure*}

\begin{figure*}[!h]
\includegraphics[width=0.49\textwidth]{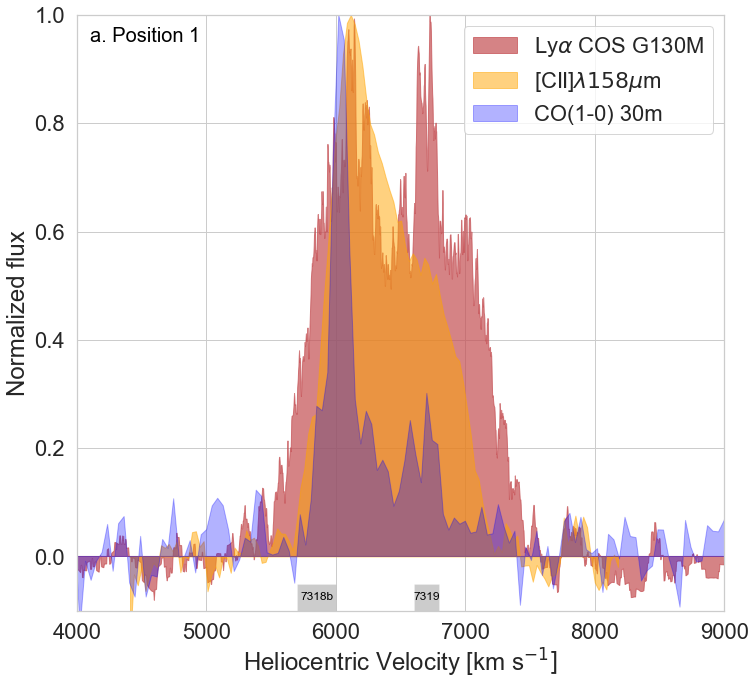}
\includegraphics[width=0.49\textwidth]{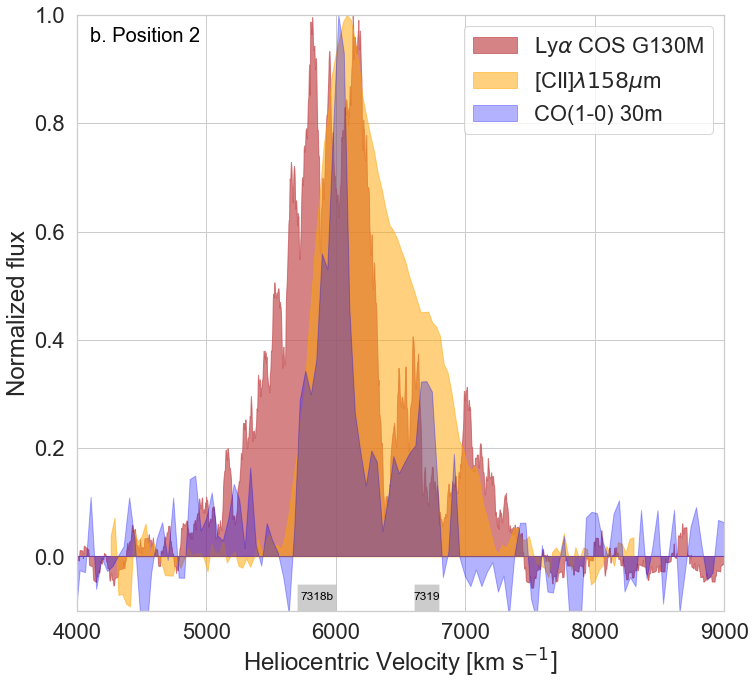}
\includegraphics[width=0.49\textwidth]{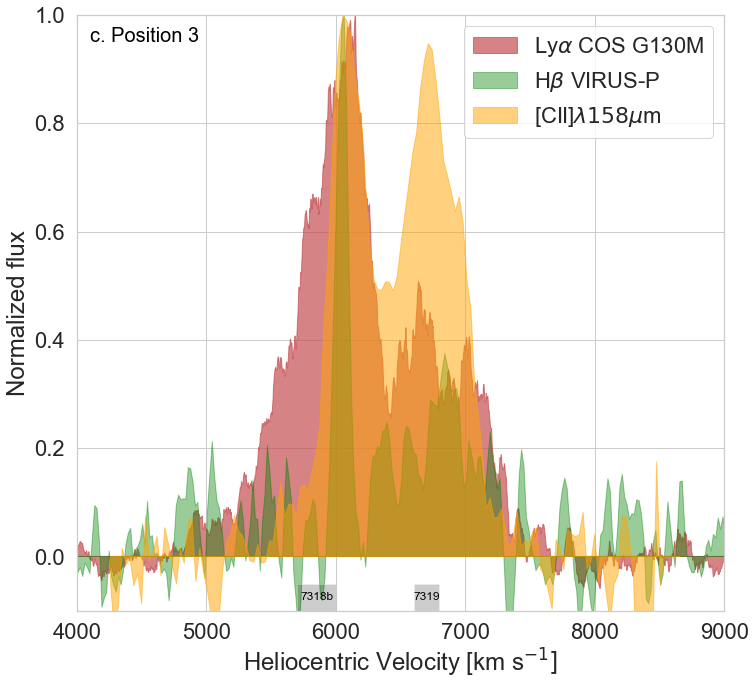}
\includegraphics[width=0.49\textwidth]{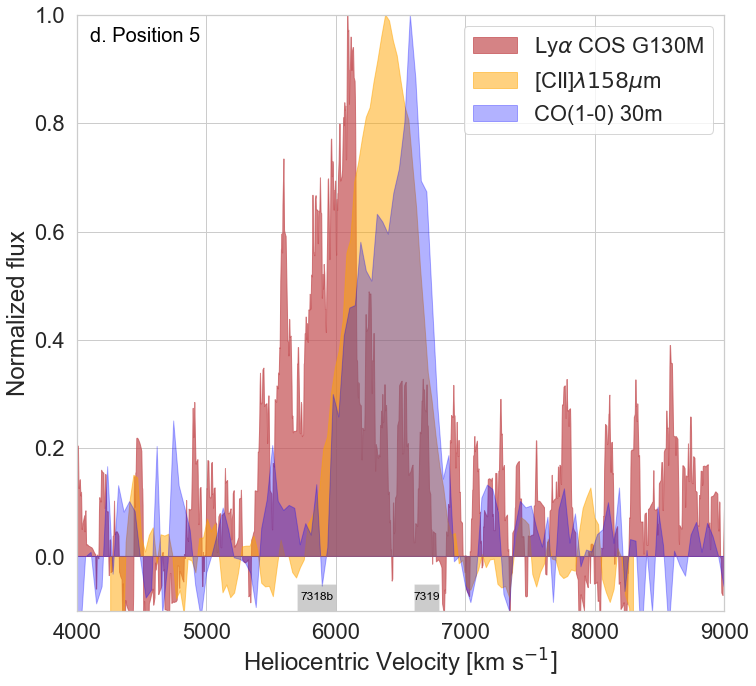}
\includegraphics[width=0.49\textwidth]{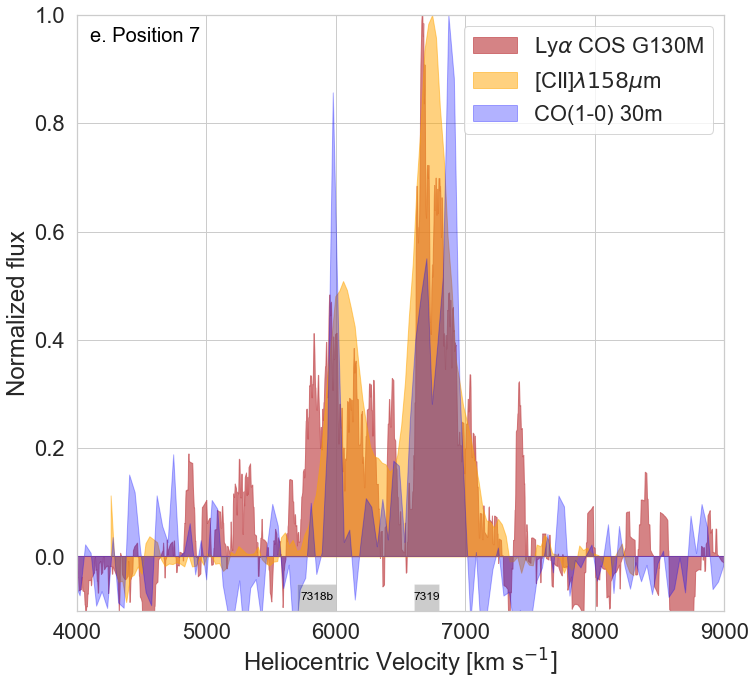}
\caption{Comparison of \Lya\ (red) , Herschel \CII\ spectra, and IRAM single-dish 30m CO(1-0) line profiles for the 5 positions observed. For position 3, where no IRAM 30m data is available, we added the comparison to the \Hb\ line taken with the GCMS (VIRUS-P) IFU Spectrograph.}
\label{fig:hcg92_COS_CII}
\end{figure*}

\section{Extremely broad \Lya\ line and kinematics of the ionized medium}

\subsection{Individual COS pointings: line fluxes and kinematics}

We detect strong and broad \Lya\ (1215.67~\AA) emission from all the five observed regions shown in Fig.~\ref{fig:SQ_contours_COS_on_WFC3Ha}. The \Lya\ line profiles for each of the 5 observed positions, shown in Fig.~\ref{fig:hcg92_COS_CO}a-e and \ref{fig:hcg92_COS_CII}a-e (in red), have complex, multi-peaked profiles. The observed coordinates and integrated \Lya\ and molecular hydrogen (H$_2$) line fluxes for each region are presented in Table~\ref{table:fluxes}. 
Our observations involve only a few COS sight-lines and therefore provide a sparse view of the \Lya\ emission across the intra-group medium. It is therefore not possible to quantify the total \Lya\ luminosity from the intra-group medium, and we will restrict our analysis to the comparison of line fluxes at the COS positions.  

The \Lya\ flux averaged over the 5 COS beams amounts to $\approx 40$\% of the warm \Hmol\ IR line emission (see last column of Table~1), which is the dominant cooling channel in the shocked intra-group medium \citep{Appleton2017} and similar to that of the \CII\ line and X-rays as well \citep[see Table~1 in][for a summary of the energy budget across gas phases]{Guillard2009}. Remarkably, on average, the \Lya\ line luminosity is comparable to that of much cooler and hotter gas, assuming that the ratio of \Lya\ to \Hmol\ emissions is the same over the whole shocked intra-group medium. Sect.~\ref{subsec:dissipation} discusses the implications of this observational result on the properties of kinetic energy dissipation in the intra-group medium of SQ. We also note that the \Lya\ flux varies by a factor of $\approx 10$, between the bridge (faintest) and the ridge 1 (brightest) positions. 

In Table~\ref{table:Lya_widths}, we gather our measurements of the widths of the \Lya\ lines for the five observed positions, as well as for the stacked spectrum. There is a large variation of the widths of the \Lya\ lines, with Full Width at Zero Intensity (FWZI) up to $\approx 2100$~\kms. This is remarkable, given that the COS beam is sampling a small region of intergalactic space between the galaxies (the 2.5~arcsec beam diameter corresponds to 1.1~kpc)\footnote{Stephan's Quintet is assumed to be at a distance of 94~Mpc}. This suggests that the COS beams likely probe the same large-scale organised structure in the filament seen at other wavelengths, rather than small individual emission regions. This becomes clear when we compare the \Lya\ profiles with \Hb\ lines, shown as red and green profiles respectively on Fig.~\ref{fig:hcg92_COS_CO}, and with \CII157.7$\mu$m lines \citep{Appleton2013}, shown as yellow profiles on Fig.~\ref{fig:hcg92_COS_CII}. In many cases, the complex \Lya\ profile shapes track approximately the main kinematic features from the other lines which are known to show large-scale coherence (see for example \citealt{Rodriguez-Baras2014,DuartePuertas2019}).  
 
Fig.~\ref{fig:hcg92_COS_CO} and \ref{fig:hcg92_COS_CII} show the varying velocity, profile shapes, line widths, and strength of the \Lya\ emission, compared with other line profiles at the same positions. The broadest \Lya\ line we observed is seen in Fig.~\ref{fig:hcg92_COS_CO}a and \ref{fig:hcg92_COS_CII}a, obtained at position HCG92-1 (see Fig.~\ref{fig:SQ_contours_COS_on_WFC3Ha}, and Table~\ref{table:Lya_widths}), near the center of the giant H$_2$ emitting filament. In this profile, the \Lya\ emission tracks quite well the shape of the \Hb\ and \CII\ profiles at low heliocentric velocity, but deviates strongly at higher velocities. The \Lya\ emission extends to at least V$_{helio}= 7600$~\kms, whereas both the \CII\ and \Hb\ emission fall almost to zero flux at velocities of no more than 7300~\kms, compared to the barycentric systemic velocity of 6600~\kms. The excess emission seen above 7300~\kms\ may be evidence of resonant scattering and bulk motions of the scattering medium, which is common in \Lya\ systems (see Sect.~\ref{subsec:broadening} for a discussion of the origin of the line broadening). This is supported by the observation that the line profiles, taken with both the Mitchell Spectrograph and {\it Herschel} generally occupy a more limited range of radial velocities compared with the \Lya\ emission, despite being taken over larger beam sampling areas than the COS data, i.e. $4 \times 4$~arcsec$^2$ for GCMS and $9.4 \times 9.4$~arcsec$^2$ for the \CII157.7$\mu$m line, see \citet{Appleton2013}. 

Other examples of possible resonant scattering wings in the \Lya\ profiles compared with the \CII\ and \Hb\ emission lines are the blue wing seen in HCG-2 (Fig.~\ref{fig:hcg92_COS_CII}b) at V$_{helio} < 5800$~\kms and the blue wing in HCG-5 (Fig.~\ref{fig:hcg92_COS_CII}d). In both cases, the \Lya\ emission extends significantly blueward of the \Hb\ line by velocities of up to 200-300~\kms. 

\subsection{Comparison with CO~(1-0) line profiles}

We compare the \Lya\ and \Hb\ lines with CO~(1-0) line profiles obtained respectively with the IRAM NOEMA interferometer and the single-dish IRAM 30m telescope \citep[from][]{Guillard2012a}, respectively in Fig.~\ref{fig:hcg92_COS_CO} and \ref{fig:hcg92_COS_CII}.
Our CO~(1-0) observations with the IRAM NOEMA interferometer, which will be presented in a companion paper (Guillard et al. in prep.), show giant, kpc-scale molecular complexes of a few $10^8\,$M$_{\odot}$ in the shock, some of them associated with \Ha-emitting regions. From the cleaned CO~(1-0) line emission map, we have extracted spectra within beams centered on the COS positions, using a synthesized beam of $4.3 \times 3.5$ arcsec$^2$, of position angle P.A.=100$^{\circ}$. We used the IRAM GILDAS mapping suite of routines\footnote{\url{https://www.iram.fr/IRAMFR/GILDAS/doc/pdf/map.pdf}} to perform the extractions, and then exported the spectra into fits files. The line parameters are gathered in Table~\ref{table:fluxes}. The CO~(1-0) line intensity was computed by integrating the line profile, and the central velocity and line velocity dispersion were computed after a Gaussian fit to the profile. The comparison between the CO~(1-0) and \Lya\ line profiles is shown in ~Fig.~\ref{fig:hcg92_COS_CO} with the caveat that the NOEMA beam is twice as large as the COS beam. It is striking that the CO lines detected with NOEMA are much narrower than the \Lya\ lines, except in the Northern star-forming region (SQ-A, position 7) where the main CO velocity component at 6700~\kms\ matches the \Lya\ profile. Those molecular complexes are much larger than the small
scale structure of the neutral gas through which \Lya\ photons scatter out of the intra-group medium. However, the spatial resolution of the NOEMA observations being 1.8~kpc, those giant molecular complexes could well break down into much smaller clumps with sometimes large shear motions between them ($\approx 100$~\kms).
We also note that there may be a more diffuse, extended molecular component, which may be filtered-out by the interferometer. This is illustrated on Fig.~\ref{fig:hcg92_COS_CII} where we compare the \Lya\ and \CII\ line profiles with CO(1-0) from single-dish IRAM 30m data. In this case, the single-dish CO line profiles are much broader than the interferometric data.

\subsection{Overview of the Large-scale motions of the intra-group medium gas}

\begin{figure*}
\includegraphics[width=0.99\textwidth]{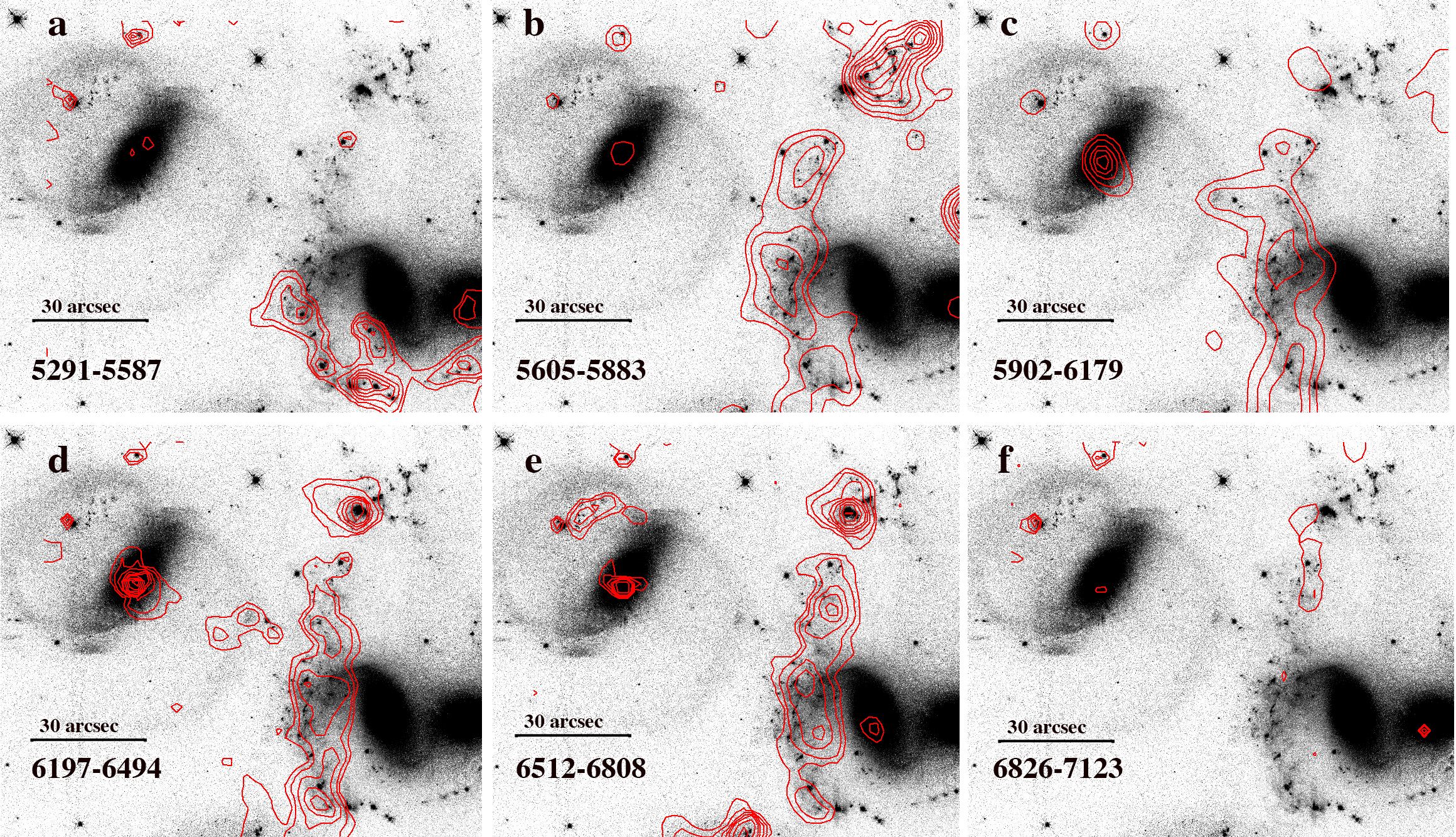}
\caption{Contours of \Hb\ surface brightness emission (in red) from Stephan's Quintet based on the Mitchell Spectrograph observations covering the full range of heliocentric velocities from 5291 to 7123~\kms\ in which emission was detected. Each sub-figure (a to f) shows the integrated surface brightness of the emission in units of $ 2.5, 5,10, 15,20, 25, 30 ,40, 50~$and$~60~\times~10^{-18}~erg~ s^{-1}~cm^{-2}~arcsec^{-2}$ over the range of velocities indicated. Subfigure (b) cover the velocity range which includes the intruder galaxy (V$_{sys}$=5774~\kms), whereas the barycenter of the group (V$_{sys}$=6600~\kms) is represented in panel (e). Gas is also seen at intermediate velocities in the bridge between the main filament and the Seyfert galaxy NGC~7319 (panels c and d). Note a possible outflow from NGC~7319. All of the contours are projected against the HST WFC3 F665N (\Ha) greyscale image of the system.}
\label{fig:VIRUS-P}
\end{figure*}

Part of the complexity of the \Lya\ line profiles can be understood when the full picture of the ionized gas (as measured optically) is explored. Other authors have presented 2-dimensional spectral maps of the optical emission lines in Stephan's Quintet \citep{Iglesias-Paramo2012, Konstantopoulos2014, Rodriguez-Baras2014, DuartePuertas2019}, but we will use our own data from the GCMS spectrograph to provide an overview and context for the observed profiles. In Fig.~\ref{fig:VIRUS-P}, we show a sequence of representative \Hb\ surface-brightness emission maps for the inner SQ group. For context, the radial velocity of the intruder galaxy, NGC~7318b is V$_{sys}$= 5774~km s$^{-1}$, and it is thought to be entering the group from behind with a blue-shifted discrepant velocity of almost 1000~\kms\ relative to the rest of the group members and the group-wide gas \citep{Xu2003,Hwang2012}. A component of the ionized gas (Fig.~\ref{fig:VIRUS-P}a and b) follows the spiral arm and \HII\ regions seen in NGC~7318b, as expected if some of the gas was part of that galaxy. Gas is also seen from the nucleus and northern disk of  NGC~7319,~(Fig.~\ref{fig:VIRUS-P}c, d and e), but with significant emission from the main North-South shocked filament. Gas at intermediate velocities (Fig.~\ref{fig:VIRUS-P}c and d) is spread along the filament, but also in the bridge between NGC~7318b and NGC~7319. The bridge is fragmentary in nature at H$\beta$, and occupies as narrower range of velocities compared with the main filament. 
The pile-up of gas with such a wide range of velocities along such a narrow structure in the main filament is a unique and remarkable feature of the Stephan's Quintet system.

\subsection{Stacked spectra and the detection of the \CIV$\lambda \! \lambda$1548,1550\AA\ doublet}

\begin{figure}
\includegraphics[width=0.49\textwidth]{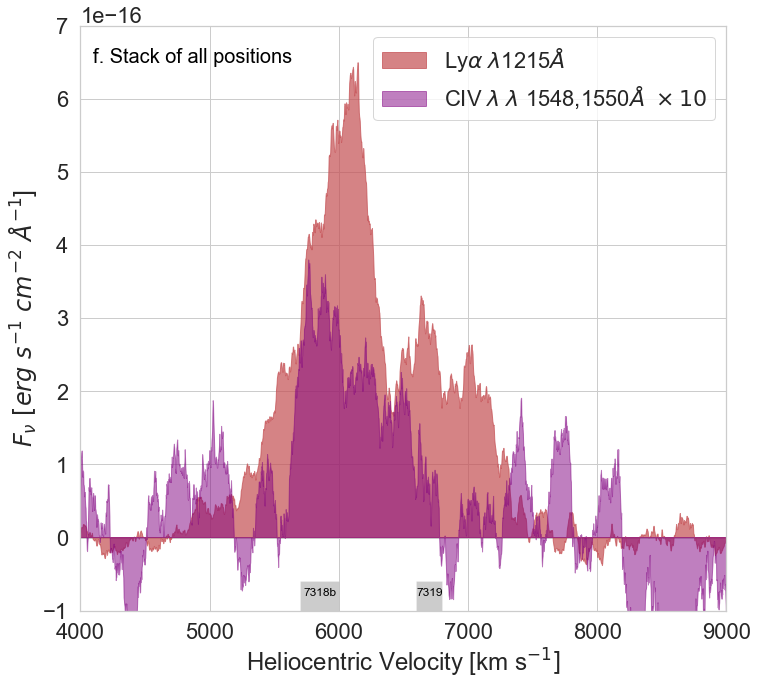}
\caption{\Lya\ line (red, grism G130M) and \CIV\ doublet (purple, grism G160M) HST COS spectra stacked over all 5 observed positions in the intra-group medium of Stephan's Quintet. The flux of the \CIV\ line has been multiplied by a factor of 10 for clarity. The widths at 20\% and 50\% of the stacked \CIV\ peak flux are, respectively,  $1100 \pm 60$ and $900 \pm 40$~\kms.}
\label{fig:hcg92_allstack}
\end{figure}

In Fig.~\ref{fig:hcg92_allstack} we show the averaged spectra of both \Lya\ and the \CIV$\lambda$1549\AA line, stacked over the five observed positions. We found all ten G160M integrations setting (two wavelengths per position) free of strong fixed pattern noise features and thus suitable for co-addition. The \CIV\ line detected on the stacked spectra is centered around the heliocentric recession velocity of the intra-group gas ($\approx$ 6000 \kms) and is very broad (FWHM $> 1000$~\kms). The profiles of the stacked \Lya\ and \CIV\ lines also show some similarities, with, for \Lya, a brighter low-velocity component around $6000-6200$~\kms and a fainter shoulder around $6300-7000$~\kms. We note that the noise in the stacked spectrum is not Gaussian, indicating that some low-level pattern noise is present, which makes the estimate of the S/N ratio of the line uncertain. The individual spectra are not very useful and show weak detections ($\approx 2\sigma$) for all five positions. 

\section{Absorption and scattering of \Lya\ photons in the intra-group medium}

By comparing the \Lya, \Hb, \CII\ and CO spectra, we can deduce at which velocities along the line of sight the \Lya\ photons are mostly absorbed. Generally speaking, looking at Fig~\ref{fig:hcg92_COS_CO} and Fig.~\ref{fig:hcg92_COS_CII}, the \Lya\  and \CII\ profiles are sometimes closer in shape than \Lya\ and \Hb. This result recalls the work of \citet{Appleton2013} showing that most of the \CII\ line emission is associated with the warm ($T > 100$~K) molecular gas (\Hmol), and that the \CII\ emission cannot be accounted for by recombination in the warm ionized medium. Therefore, both \CII\ and \Lya\ trace gas heated by mechanical energy dissipation.

There are four possible examples of where \Lya\ absorption is taking place. In Fig.~\ref{fig:hcg92_COS_CII}b and c we see that the high velocity component of the double profile seen in both \Hb\ and \CII\ is significantly suppressed compared with the low-velocity component. For Fig.~\ref{fig:hcg92_COS_CO}b in particular, the peak in the \Hb\ profile falls close to a strong dip in the \Lya\ profile, which appears
as two small wings on either side of the dip.  For Fig.~\ref{fig:hcg92_COS_CII}c, the feature seen in both \CII\ and \Hb\ is largely suppressed above 6500~\kms. Also, in Fig.~\ref{fig:hcg92_COS_CO}d and \ref{fig:hcg92_COS_CII}d, the \Lya\ profile
is shifted blueward of the main \CII\ and \Hb\ peaks, suggestive of absorbing gas centered at V$_{helio}$ = 6500-6700~\kms. Asymmetric \Lya\ profiles like this are often associated with radial outflows in galaxies \citep[e.g.][]{Heckman2011}. The COS pointing (HCG92-5) samples the gas in the so-called "AGN bridge" \citep{Cluver2010}, a linear H$_2$ filament that is apparently separate from the main collisional shock in SQ. A strong shear in the velocity field of the \CII\ emission was noted by \citet{Appleton2013} in that region. 
The spectrum of the CO~(1-0) emission from that direction (see Fig.~\ref{fig:hcg92_COS_CO}d and Fig.~\ref{fig:hcg92_COS_CII}d), shows significant CO emission at the high velocity side of the profile, which would be consistent with absorption. Finally, HCG92-7 shows, in Fig.~\ref{fig:hcg92_COS_CII}e, the opposite effect. In this case, again considering the high-velocity component of the double-horned profile only, we see that the \Lya\ is significantly redshifted with respect to the \Hb\ emission, with a sharp drop in emission as one approaches the peak of the \Hb\ (around V$_{helio}$ = 6600-6700 \kms). This may be another example of asymmetric absorption, with resonant scattering to the red-side of the wings of the kinematics. In summary, we see that two regions in the main emission-line filament are almost free of absorption, whereas other regions show strong absorption. Even in those cases, at least some of the \Lya\ emission is able to resonantly scatter, and undergo many scatterings within the gas before eventually escaping into the wings of the velocity profile where the optical depth is much lower. In Sect.~\ref{subsec:broadening} we estimate the escape fraction of the \Lya\ photons and the number of scatterings, and we discuss the multi-phase structure of the intra-group medium gas. We also remind that the differences between the \Lya\ and \Hb\ spectral profiles may not only be due to \Lya\ scattering, because collisional excitation could also contribute.

Does this interpretation make sense in terms of the expected line ratios for the hydrogen lines? 
In Table~~\ref{table:fluxes} we also present the \Hb\ line fluxes and \Lya\ to \Hb\ flux ratios integrated over the 5 different sets of spectra.  It is interesting that HCG92-1 and 2 both show ratios (31 and 22 respectively) consistent with little or no absorption when compared with Case B recombination (Case B predicts for T = 10$^4$ K and typical interstellar densities  a flux ratio F(\Lya)/F(\Hb) of $\sim$ 33). The lack of extinction inferred from the line ratio for HCG-1 is entirely consistent with our previous description of the close similarity between the \Lya\ and \Hb\ line profile shapes.  The slightly lower \Lya\ to \Hb\ ratio is consistent with the increased absorption in the red component of the \Lya\ profile.  For both HCG92-3 and 5, F(\Lya)/F(\Hb) are significantly lower than Case B, suggesting stronger absorption, which is again consistent with the line profiles. Finally,  HCG92-7 (which is associated with
the extragalactic H{\sc ii} regions and contains significant star formation and dust), shows the highest deviation from Case B (F(\Lya)/F(\Hb) $\sim$ 2) suggesting that, at that position, most of the \Lya\ photons are absorbed by dust.

\begin{figure*}
\includegraphics[width=0.5\textwidth]{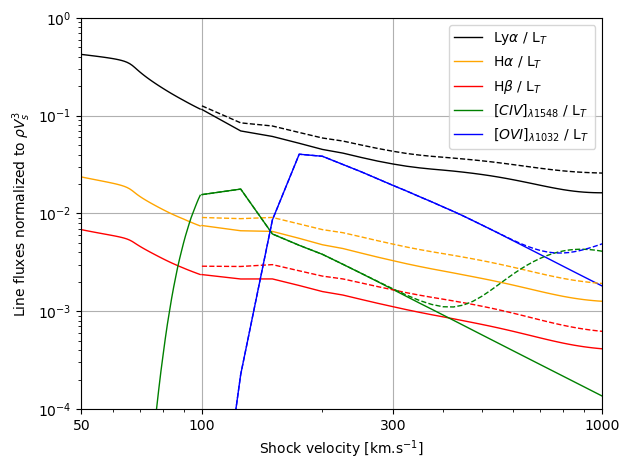}
\includegraphics[width=0.5\textwidth]{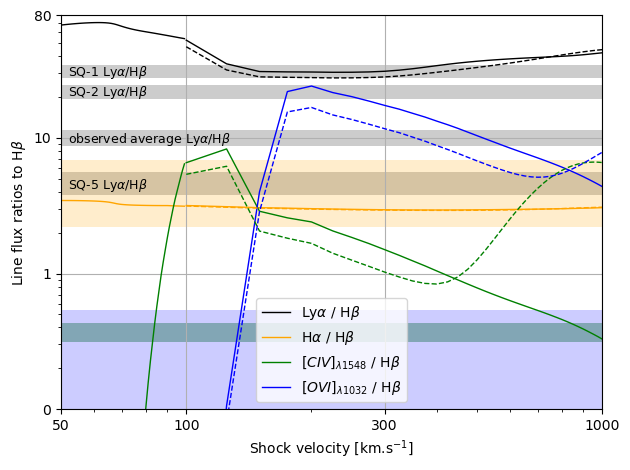}
\caption{Results from the MAPPINGS V shock models library from \citet{Alarie2019}, covering a range in velocities of 50-1000~\kms. Solid lines show the contribution of the shock only and dashed lines include the contribution of the radiative precursor at velocities $V_s > 100$~\kms. In this grid of models, the preshock Hydrogen density is \nH = 10~H~cm$^{-3}$ and the pre-shock magnetic field intensity is $B = 1\,\mu$G. \textit{Left:} Line fluxes normalized to $\mu m_H n V_s^3$ as a function of  the shock velocity $V_s$, $\mu =1.3$ is the mean molecular weight for neutral gas, and \mH\ the atomic mass unit. For weakly-magnetized shocks like those presented here, $\rho V_s^3$ is close to the total radiative flux of the shock. \textit{Right:} Line emissivity ratios to \Hb\ flux versus shock velocities, with the same chemical abundances (Solar) and shock parameters as \citet{Allen2008}. The grey bands show \Lya/\Hb\ ratios for the three positions (1, 2 and 5) associated to broad optical line emission and shock excitation, and the value averaged over all positions (see~Table~\ref{table:optical_lines}). The orange and green bands show respectively the range of observed \Ha/\Hb\ ratios and the stacked \CIV/\Hb\ value with a $\pm 20$~\% uncertainty. The blue band show the $3\sigma$ upper limit on the \OVI/\Hb\ ratio.
}
\label{fig:UV_lines_shock_models}
\end{figure*}

\section{Origin and properties of the \Lya\ emission in the intergalactic medium of SQ}
\label{sec:discussion}

\subsection{Shocks and turbulent mixing layers}
\label{subsec:shocks_mixing}

Strong \Lya\ line emission, with typical FWHM of 200-1500~\kms, is often observed in high redshift galaxies \citep[e.g.][]{Tapken2007}, where \Lya\ photons produced by powerful starbursts scatter off neutral gas carried in outflows. Similar broad \Lya\ profiles are sometimes seen in the inner regions of \Lya\ nebulae associated with luminous high-z quasars \citep[e.g.][]{Ginolfi2018}. However, these extreme conditions are the antithesis of those seen in the SQ filament, where the star formation activity is very weak and, globally, the \Lya\ emission is mainly powered by dissipation of mechanical energy. In this section, we argue that both radiative shocks and turbulent mixing layers may contribute to powering the observed \Lya\ line emission. 

Shocks having velocities high enough to reach temperatures capable of collisionally exciting electronic states of atomic Hydrogen (above $10^4$~K) are a strong source of \Lya\  photons \citep[e.g.][]{Shull1979, Dopita1996, Lehmann2020}. Due to collisional ionization of hydrogen,  \Lya\ and \Hb\  photons are mostly produced in gas at temperatures smaller than T$=10^5\,$K, with collisional excitation dominating recombination  for T$>10^4\,$K \citep{Raga2015}. In the intra-group medium of SQ, it is likely that a wide distribution of shock velocities is present \citep{Guillard2009}. In this paper we do not attempt at a detailed modelling of the line emission from shocks. We rather aim at qualitatively determining which shocks contribute the most to the \Lya\ and \CIV\ line emission. 

To do so, we use the results from the MAPPINGS~V shock code library. The physics of the models is fully described in \citet{Sutherland2017} and the data in \citet{Alarie2019}. These shock models are based on \citet{Allen2008}, but the new models extend the predictions for shocks and radiative precursors to shock velocities smaller than 100~\kms, as well as up to 1500~\kms. These models include expanded atomic cooling lines and comprise a wide range of shock precursor conditions, from completely neutral gas through partially ionized and fully ionized. Magnetic fields are also included as they can strongly impact the compression and temperature of the post shocked gas. 
Figure~\ref{fig:UV_lines_shock_models} shows results from the MAPPINGS~V shock models for a neutral atomic pre-shock gas with density \nH = 10~H~cm$^{-3}$ and a pre-shock magnetic field intensity of $B = 1\,\mu$G. 
The pre-shock gas ionisation fraction is computed from the UV emission generated by the shock. We only show models for shock velocities above $V_s > 50$~\kms\ because modelling of H$_2$ line emission shows that, in SQ, lower velocity shocks are molecular shocks \citep{Guillard2009}.

The left panel of Figure~\ref{fig:UV_lines_shock_models} shows some optical and UV lines fluxes normalized\footnote{See the energy conservation equation Eq.~4 in \citep{Allen2008}. For negligible magnetic energy, the total radiative flux of the shock is  $F_{tot} = 0.5 \rho V_s^3$. 
} to $\rho V_s^3$, i.e. the sum of the kinetic and enthalpy fluxes, where $\rho = 1.4$ \nH \mH\  is the gas mass density, as a function of the shock velocity $V_s$. The solid lines show the fractional line luminosities from the shock only, while dashed lines include the contribution of the precursor for shock velocities $V_s > 100$~\kms. 
Due to collisional ionization of Hydrogen atoms, the fraction of the shock emission accounted for by \Lya\ photons is the highest for shock velocities smaller than 100~\kms\,  for which Hydrogen excitation is mainly collisional. Faster shocks do contribute to \Lya\ emission but to a lesser fraction of the total radiated power. For shock velocities typically above 150~\kms, the \Lya\ shock emission comes mainly from the photo-ionisation of the post-shock gas that has cooled down to $\sim 10^4\,$K. The UV emission produced in fast shocks is also in part processed into \Lya\ photons by hydrogen photo-ionization in the radiative precursor \citep[][]{Sutherland2017}. The \CIV\ and \OVI\ line fractions peak at shock velocities higher than for \Lya, 120 and 200~\kms\ respectively. The contribution of photo-ionization to the \Lya\ emission could thus be significant in the intra-group medium of SQ. This contribution would still be associated with the dissipation of mechanical energy since in SQ the UV flux is mostly produced by  shocks. 

In the right panel of Fig.~\ref{fig:UV_lines_shock_models} we show model predictions for strength of various emission line fluxes normalized to the \Hb\ line, and we compare them to the observed values, averaged for the 5 positions. 
The \Ha/\Hb\ ratio in the models is always above the case B value (2.86), and slightly below the stacked value (\Ha/\Hb = 4.3), although observations towards the 5 positions span a large range, i.e. \Ha/\Hb = 2.2 -- 6.9.  
The rise of the \Lya/\Hb\ at low velocities ($V_s < 150$\kms) is due to collisional excitation in the post shock gas, since the precursor gas entering the shock is neutral \citep{Sutherland2017}. 
We also note that both the low \CIV/\Hb\ observed ratio and the \OVI/\Hb\ upper limit\footnote{computed with a $3\sigma$ upper limit on the \OVI\ line flux measured on the G130M stacked spectrum.} point to shocks below 150~\kms.   

Another contribution to the generation of \Lya\ emission could be irradiated molecular shocks at lower velocities than those presented in Fig.~\ref{fig:UV_lines_shock_models}. Shocks at velocities 30--50~\kms driven into molecular gas at typical densities \nH$ = 10^4$~cm$^{-3}$ produce a strong \Lya\ radiation \citep{Lehmann2020}. However, the \Lya / $L_T$  ratio in such molecular shocks is lower (10--20\%) than for the shocks in atomic gas at similar velocities presented in Fig.~\ref{fig:UV_lines_shock_models}.  
In addition to shocks, turbulent mixing of gas phases is likely to contribute to the emission of UV line emission \citep[e.g.,][]{Slavin1993, Kwak2010} and to the overall energy dissipation in the multiphase medium of SQ. Turbulent mixing layers may also explain the high \CIV / \OVI\ line flux ratio ($>0.7$), akin to what is observed in the circum-galactic medium \citep[e.g.][]{Shull1994, Fox2011}. 
 
\subsection{Scattering of \Lya\ photons in a highly clumpy medium with large bulk motions as a source of line broadening} 
\label{subsec:broadening}

\citet{Neufeld1991} and \citet{Charlot1993} first emphasized the impact of clumping and the multi-phase nature of astrophysical media on the \Lya\ line strength and spectral shape. This paper does not attempt  at a quantitative modelling of the \Lya\ emission in the multiphase intra-group medium of SQ. We only present  some qualitative suggestions within the framework of \Lya\ radiative transfer in a clumpy medium \citep{Zheng2002, Verhamme2015, Gronke2016, Gronke2017}.

 mean ratio between \Lya\ and {H$\beta$} fluxes measured on stacked spectra $\approx 10$ (Table~\ref{table:fluxes}) may be compared with the intrinsic value $\approx 30$ for hydrogen recombination at temperatures of a few $10^4\,$K and $\approx 80$ for collisional excitation in the lowest velocity shocks ($ V_s < 100 \kms\ $) in the right plot of Fig.~\ref{fig:UV_lines_shock_models}. From this comparison, we estimate the escape fraction of \Lya\ photons to $\sim 10 - 30\%$. This is an effective value that varies significantly among regions and within regions with gas velocity.

In an interacting system like SQ, coherent gas flows within the SQ intra-group medium are likely to contribute to the broadening of the \Lya\ line profile. In particular, it is likely that the prominent blue-shifted scattering wings observed at Positions 2, 3 and 5 are the result of systematic velocity gradients related to the 3D geometry of the collision between the intruder and the intra-group medium. More generally, the width of the \Lya\ line does not provide a direct constraint on the number of scatterings and the gas clumping. Further, existing models of \Lya\ radiative transfer based on micro-turbulence (pure random motions) make simplifying assumptions that probably do not apply to SQ. With this caveat in mind, we provide here some indicative numbers based on these models. 

In a static medium, the number of scatterings $N_{sc}$ can be estimated from the observed frequency shift $\Delta \nu = \nu - \nu_0$, where $\nu$ and $\nu_0$ are the observed and rest frequencies of the \Lya\ line, as the following:  $N_{sc}^{1/2} = \Delta \nu / \Delta \nu _D \approx (a \tau_0)^{1/3}$, where $\Delta \nu _D$ is the thermal Doppler broadening, $a$ the damping parameter, and $\tau_0$ the line-center optical depth \citep[see][]{Neufeld1990}. Assuming $T=10^4$~K and a typical velocity shift of 100~\kms, we find $N_{sc} \approx 60$. In a multiphase medium, the \Lya\ escape fraction $f_{esc}$ depends on the dust optical depth of the clumps, $\tau _d$, and the covering factor $f_c$, i.e. the average number of clumps  along the sightline \citep{Neufeld1991}. Both the modelling of the dust emission in SQ \citep{Natale2010, Guillard2010} and studies of the molecular gas content \citep{Guillard2012a,Appleton2017} converge to an average column density of $N_H \approx 2 \times 10^{20}$~cm$^{-2}$ in the ridge, which translates into $\tau _d \approx 0.5$ at $\lambda = 1216$~\AA. Following \citet{Hansen2005} and \citet{Gronke2017}, we find a covering factor $f_c \approx 15$ for a fiducial escape fraction $f_{esc} = 0.15$, and $N_{sc} \approx 80$. To test these estimates, more realistic radiative transfer models including coherent gas velocity gradients would be needed, as well as very high spatial resolution observations to confirm their presence.

In conclusion, this high escape fraction, combined with the spectral evidence of \Lya\ scattering, reflects the clumpy picture that has emerged from the analysis of SQ observations, mainly the spatial correlation between the tracers of the hot, warm and cold intra-group medium phases, and the modelling of the SQ dust emission \citep{Guillard2010}.
The neutral gas (dominated by dusty molecular gas in the ridge) is in clumps 
with a high velocity dispersion and large velocity gradients. 
The clumps are embedded in a hot X-ray emitting, dust-free, plasma. 
Within such a clumpy medium, \Lya\ photons may escape through multiple scatterings off the clump surfaces \citep{Neufeld1991, Gronke2017} without being absorbed in the inter-clump dust-free plasma. The clumps must fill a small fraction of the volume  but their surface filling factor must be close to unity with multiple clumps along a given line of sight.  Further work is needed to model these observations and to assess whether the differences in spectral shapes observed for the five pointings could be accounted for by variations in the total column of dusty neutral (molecular) Hydrogen, the number of clumps along the line of sight and their kinematics.  Note that not only scattering effects but also collisional excitation could also contribute to explain the differences between the \Lya\ and \Hb\ spectra.

\subsection{A constant dissipation rate across many orders of magnitude in gas temperatures: a signpost of a turbulent cascade?}
\label{subsec:dissipation}

The observations presented in this paper brings another piece to the cooling budget puzzle of the SQ shocked intra-group medium. Putting together multi-wavelength line spectroscopy allows us to combine radiative tracers which spans a wide range of  gas temperatures, from $\gtrsim 100$~K for rotational \Hmol\ and \CII\ lines, to $5\times 10^6$~K for X-rays. Remarkably, over more than four orders of magnitude in temperature, the powers radiated by the multi-phase intra-group medium in X-rays, \Lya, \Hmol, \CII\ are comparable within a factor of a few \citep[see also Table~1 in][for a summary of the energy budget across gas phases]{Guillard2009}. This indicates that the dissipation of the kinetic energy in the SQ galaxy-wide collision involves all gas phases. Dissipation could proceed through shocks with a wide distribution of velocities and involve turbulent mixing layers. Shocks and mixing layers can be the combined result of the turbulent energy cascade within the multiphase intra-group medium. While a specific probability distribution function of shock velocities is required to make the cooling rate independent of the gas temperature, models suggest that this may be a generic property of energy dissipation in turbulent mixing layers \citep{Ji2018}.

\section{Similarities with the Circum-Galactic Media of distant galaxies}

The nature of the emission from the intra-group medium of Stephan's Quintet may have implications for our understanding of the observations and nature of the Circum-Galactic Medium (CGM) surrounding galaxies in the distant universe. 
Superficially, there are many similarities between the CGM of galaxies and the intra-group medium in SQ. The CGM of distant galaxies have evidence for: \textit{(1)} bright \Lya\ emission with broad lines ($\sigma \approx 100-500$ \kms) and line profiles which range from simple Gaussians, double-horned, to those with strong asymmetries \citep[e.g.,][]{leclercq17, vernet17, leiber18, osullivan20}; \textit{(2)} emission at a wide range of frequencies implying that the gas in the CGM is multiphase and contains cold gas \citep[e.g.][]{emonts16, emonts18, emonts19, falkendal21}; and \textit{(3)} line ratios of the UV and optical emission lines that are about the values expected for recombining, clumpy gas \citep{leiber18, cantalupo19, marino19}. The intra-group medium of SQ is also somewhat reminiscent of the very broad, shock-powered \Lya\ and CH$^+$ emission detected outside galaxies in the distant galaxy group SMMJ02399 at $z=2.8$ \citep{VidalGarcia2021}.
In all of these ways, there are similarities with the selected regions of the intra-group medium of SQ we have studied.

Of particular interest is the comparison of the \Ha\ and \Lya\ line ratios and surface brightnesses between the intra-group medium of SQ and the ICM of the Slug Nebula and MRC\,1138-262 \citep{leiber18, shimakawa18}. In the Slug Nebula and the medium surrounding the radio galaxy, MRC\,1138-262, the \Lya\ to \Ha\ line ratios are about 6 and a few respectively. These ratios are quite similar to the range of values spanned in the spectra of the regions of the SQ, less than 1 to almost 10. It is also worth noting that in the case of the Slug Nebula, the \Ha\ line is consistent with being much narrower than the \Lya\ emission over the same region \citep{leiber18} and similar to our findings. These results suggest that a fraction of the \Lya\ escapes the nebulae and that scattering plays an important role in shaping the line profiles. The surface brightness of the hydrogen lines is also quite different in these objects. In both the Slug and MRC\,1138-262, the surface brightness of \Ha\ is approximately 1-2 orders of magnitude higher than in the emission line regions we have observed in SQ. The Slug Nebula has a lower surface brightness than MRC\,1138-262 but was also estimated at a much larger distant from the QSO than that used for MRC\,1138-262 \citep[estimated at the faintest surface brightness levels in the narrow band imaging data; ][]{shimakawa18}.

Since both the Slug and MRC\,1138-262 have been observed in CO transitions, we can also compare their molecular gas surface densities. Over roughly the same regions are those used to estimate the \Ha\ surface brightnesses, the Slug has upper limits to its H$_2$ gas mass surface densities, $<$12-25 \Msun\ pc$^{-2}$ \citep{decarli21}, while MRC\,1138-262 is about 35 \Msun\ pc$^{-2}$ \citep{emonts16}. The mass surface densities of H$_2$ for SQ range from about 10-100 \Msun\ pc$^{-2}$. It is interesting that the molecular gas surface densities of MRC\,1138-262 are similar while the \Ha\ (and given the similar ratios, the \Lya) surface brightnesses are so different. This can simply be explained within the context of our analysis that both objects have strong energy injection into their circumgalactic media (a collision and a high power radio jet) which drives a turbulent cascade, but MRC\,1138-262 hosts a power, UV luminous AGN which powers its optical emission line gas. The emission from the Slug nebula is consistent with that expected from photoionized gas \citep{decarli21} while clearly this is not the case for either the regions of SQ and MRC\,1138-262. The injection of mechanical energy into their circumgalactic media plays an important role in shaping what we observe, especially in creating and sustaining dense molecular gas.

\section{Conclusions} 

We have used the COS spectrograph on HST to observe \Lya\ emission from the intergalactic gas in SQ. The observations sample five positions across the 30~kpc-wide shock. The HST data is compared with CO, \CII\ and \Hb\ spectra. We summarize the main observational results and outline our interpretation of the data.

 We detect extremely wide \Lya\ lines with a full width at zero intensity of $\approx 2000$~\kms, which exceeds the velocity range of CO, \CII\ and \Hb\ line emission. After stacking of the five HST spectra, we also detect the \CIV\ doublet. We observe significant variations in the  \Lya\ / \Hb\ spectral ratio between positions and velocity components. From the mean line ratio averaged over positions and velocities, we estimate the mean escape fraction of \Lya\ photons to be $\sim 10-30\%$. The \Lya\ lines are systematically broader than the \Hb\ ones at the same positions, which we consider as observational evidence for scattering of \Lya\ photons by the SQ intra-group medium. The difference in velocity spread is asymmetrical and amounts to $\approx 300$~\kms\ for the blue-shifted \Lya\ wings observed at three of the five positions.  

The observations provide insight on the structure of the multiphase intra-group medium in SQ. The high \Lya\ escape fraction and scattering reflect the clumpy picture suggested by the spatial correlation between the tracers of the hot, warm and cold phases of the SQ intra-group medium.  The neutral, mainly molecular, gas is in clumps embedded in the X-ray emitting, hot and dust-free plasma. \Lya\ photons must escape through multiple scatterings off the clumps.  Scattering indicates that the intra-group medium is not porous to \Lya\ photons, i.e. the neutral gas surface filling factor must be close to unity with multiple clumps along a given line of sight. A quantitative comparison with \Lya\ radiative transfer models is beyond the scope of this observational paper, but these data suggest that coherent gas flows within the SQ intra-group medium contribute to the broadening of the \Lya\ line profile.
In particular, it is likely that the blue-shifted scattering wings follow from systematic velocity gradients related to the 3D geometry of the collision between the intruder and the SQ intra-group medium. 

The bulk of the \Lya\ emission must be powered by dissipation of mechanical energy because the SQ star formation rate is small and the gas velocities span an exceptionally large range. This conclusion is in line with optical line ratios measured at our COS pointings.
\Lya\ photons are emitted by gas at temperatures smaller than the thermal energy threshold for collisional ionization ($T < 10^5\,$K). It is likely that both  
collisional excitation and recombination of photo-ionized Hydrogen contribute to the observed emission.
Due to collisional ionization of hydrogen atoms, the fraction of the shock emission accounted for by \Lya\ photons is the highest for shock velocities smaller than 100~\kms. Faster shocks do contribute to \Lya\ emission but to a lesser fraction of the total radiated power. The UV emission produced in fast shocks is in part processed into \Lya\ photons in the post-shock and the pre-shock gas.  This contribution of photo-ionized gas to the \Lya\ emission, which is also associated with dissipation of mechanical energy, could be significant.

The HST observations complement our view at the energetics of the galaxy-wide shock created by the collision of high-speed intruder galaxy with the SQ intra-group medium.
The total power emitted in the \Lya\ line is comparable to that of much cooler gas in the mid-IR rotational \Hmol and the \CII\ fine structure lines. The energy radiated in \CII,  \Hmol, \Lya\ and X-rays represents cooling from gas spanning four order of magnitudes in temperature from 100 to $10^6\,$K. The observed fluxes are comparable within a factor of a few, which indicates that roughly the same fraction of energy is dissipated per logarithmic bin of temperature. This is a remarkable result that constrains models of the turbulent energy cascade in SQ. It emphasises the possible contribution from turbulent mixing layers to  energy dissipation.  

Following the trail of mechanical energy dissipation and gas kinematic in the turbulent gas on scales smaller than $\approx 1$~kpc (the COS aperture scale) will be possible with observations of warm molecular and ionized gas with the {\it James Web Space Telescope}, and future UV-optimized telescopes. Such observations will also to look for coherent anisotropic gas flows, which are necessary in order to explain the blue scattering wings in the \Lya\ profiles seen in some of the observed positions.   

\acknowledgments
PG thank the Centre National d'Etudes Spatiales (CNES), the University Pierre and Marie Curie, and the "Programme National de Cosmologie and Galaxies" (PNCG) and the "Physique Chimie du Milieu Interstellaire" (PCMI) programs of CNRS/INSU for there financial supports. We thank Daniel Kunth and Brigitte Rocca for very useful physical and technical discussions about \Lya\ scattering. Support for Program number HST-GO-13321.001-A was provided by NASA through a grant from the Space Telescope Science Institute, which is
operated by the Association of Universities for Research in Astronomy,
Incorporated, under NASA contract NAS5-26555. PA thanks Guillermo Blanc (Carnegie Observatories) and Emily Freeland (formerly Texas A\&M University) for providing assistance with observations, and software/data reduction associated with the Mitchell Spectrograph. This work used observations carried out under project number U020 (P.I. Guillard) with the IRAM NOEMA Interferometer, reduced and analysed with the GILDAS software (\url{https://www.iram.fr/IRAMFR/GILDAS}). IRAM is supported by INSU/CNRS (France), MPG (Germany) and IGN (Spain). We also thank the referee who has contributed to improve both the figures and content of the paper.

\bibliography{Guillard_HST_COS_SQ}

\end{document}